\documentstyle[aps,epsf]{revtex}  

\begin{document}        
\newcommand{\aelec}{$A_{e}$}
\newcommand{\abquark}{$A_{b}$}
\baselineskip 14pt
\title{Measurements of $A_{LR}$, $A_{lepton}$ and $A_{b}$ from SLD}
\author{Jorge P. Fernandez}
\address{Department of Physics, University of California, Santa Cruz,
CA 94064}
%
\maketitle              

\begin{abstract}        
     We present the measurements of the leptonic asymmetries in $Z^0$
decays measured by the SLD experiment at SLAC. A data sample of
approximately 550,000 $Z^0$ bosons are used for the preliminary
measurement of $A_{LR}(A_{e})$ and the leptonic final state
measurements of $A_{e}$, $A_{\mu}$ and  $A_{\tau}$, representing the
entire set of 1992-1998 SLD runs. When combining all results, a
preliminary value for the effective weak mixing angle is obtained:
${\sin}^2{\theta}_{W}^{eff} = 0.23109 \pm 0.00029$. We also present the
direct measurements of $A_{b}$ from left-right forward-backward
asymmetries, using partial data samples from 1993-1998 SLD runs. The
preliminary value combines jet charge, cascade kaon and semileptonic
decay analyses to obtain $A_{b} = 0.866 \pm 0.036$.
\
\end{abstract}   	

\section{Introduction}               

The SLC linear electron-positron collider\cite{SLC} has long been a powerful
facility for testing the Standard Model via measurements of
electroweak couplings at the $Z^0$ pole. Its unique attributes of a
highly longitudinally polarized electron beam ($P_{e} \approx 75$\%) and
small spot size (1.5 x 0.8 x 700) $\mu$m$^3$ in (x,y,z) have recently
been augmented by improved luminosity for the 1997-1998 SLD run and
these have proved advantageous for precision electroweak
measurements. The SLD detector\cite{SLD} is designed to take advantage of these
special attributes of the SLC.

\subsection{Asymmetries}
The polarized differential cross section at Born level for the process,
$e^{+}e^{-} \rightarrow Z^{0} \rightarrow f\bar{f}$ is given by:
\begin{eqnarray}
{{\rm d}\sigma^{f} \over {{\rm d}z}} \propto \left( 1-P_{e}A_{e}
\right)\left( 1+z^{2} \right)+2A_{f}\left(A_{e}-P_{e} \right)z
\label{eqn:PolCross} 
\end{eqnarray}
where $A_{f} = 2v_{f}a_{f}/(v_{f}^{2}+a_{f}^{2})$ is the final state
coupling, $P_{e}$ is the electron beam polarization ($\equiv +1$ for
right-handed beams), $z = \cos{\theta}$ where $\theta$ is the angle of
the final state fermion with respect to the electron beam axis, and $v_{f}$ and
$a_{f}$ are the vector and axial vector couplings which specify the
Z-f coupling.

One can define the left right asymmetry which equals the initial state
coupling asymmetry $A_{e}$,
\begin{eqnarray}
{A_{LR}^{0}} & \equiv & {1 \over \left|{P_{e}}\right|}{ {\sigma
\left(e^{+}e_{L}^{-} \rightarrow Z^{0} \right) - \sigma
\left(e^{+}e_{R}^{-} \rightarrow Z^{0} \right)} \over {\sigma
\left(e^{+}e_{L}^{-} \rightarrow Z^{0} \right) + \sigma
\left(e^{+}e_{R}^{-} \rightarrow Z^{0} \right)} } 
= {1 \over {P_{e}}}{\left[ P_{e} { {2v_{e}a_{e}} \over {
v_{e}^{2}+a_{e}^{2}}}\right]} = A_{e}
\label{eqn:Alr} 
\end{eqnarray}
Although one can use all final states to build the asymmetry, in
practice, our cuts select hadronic final states: leptonic final states
are treated separately (see section IV). If $P_{e}$ can be measured precisely, then $A_{LR}$ provides us with an effective way to measure ${\sin}^{2}{\theta}_{W}$:
\begin{eqnarray}
{A_{LR}^{0}}& = & { {2 \left(1-4{\sin}^{2}{\theta}_{W}^{eff} \right)} \over 
{1+ \left(1-4{\sin}^{2}{\theta}_{W}^{eff} \right)^{2}} }
\label{eqn:AlrEff} 
\end{eqnarray}
where the evolution of ${\sin}^{2}{\theta}_{W} \rightarrow
{\sin}^{2}{\theta}_{W}^{eff}$ signifies the convention of
making Eqn.~\ref{eqn:AlrEff}  the {\it definition} of the weak mixing angle.
Thus higher order effects must be explicitly accounted for when
comparing this value to non-$Z^{0}$-pole measurements.

Additionally, one can construct a forward-backward left-right
asymmetry and extract the final state coupling $A_{f}$ ,

\begin{eqnarray}
{\tilde A}_{FB}^f(z) & = &
{[\sigma^f_L(z) - \sigma_L^f(-z)] - [\sigma_R^f(z) - \sigma_R^f(-z)] \over
  \sigma^f_L(z) + \sigma_L^f(-z)  +  \sigma_R^f(z) + \sigma_R^f(-z)  }
\nonumber \\
& = & |P_e|A_f {2z \over 1 + z^2}.
\label{eqn:Alrfb} 
\end{eqnarray}

For b-quark final state fermions and assuming full detector
acceptance, Eqn.~\ref{eqn:Alrfb} yields ${\tilde A}_{FB}^b =
(3/4)|P_e|A_b$. It is interesting to note that while $A_{LR}$ is very
sensitive to the weak mixing angle, $A_b$ in insensitive to
${\sin}^{2}{\theta}_{W}^{eff}$, and instead is sensitive to vertex
effects which in the Standard Model are expected to be negligible at the
$Z^{0}e^{+}e^{-}$ vertex.

\section{Polarization Measurements}

Precision polarimetry of the SLC electron beam is accomplished
with the Compton Polarimeter~\cite{compton}, which employs Compton scattering
between the high-energy electron beam and a polarized
Nd:YAG laser beam ($\lambda = 532$nm) to probe the
electron beam polarization. The Compton scattered electrons, 
which lose energy in the scattering process
but emerge essentially undeflected, are momentum analyzed by the
Compton spectrometer downstream of the SLD interaction point,
beyond which they exit the beam-line vacuum through a thin window,
and enter a threshold \v{C}erenkov detector segmented transverse to the
beam-line.

A history of polarization performance is given
elsewhere~\cite{compton2}. The preliminary measurement for 1998 is $P_e
= 73.1 \pm 1.01$(syst)\% where the relative systematic uncertainty can be
expected to improve substantially (to below the 0.7\% level) when the analysis is complete.

\section{Measuring $A_{LR}$}

$A_{LR}$ is essentially measured with hadronic final states
only, as the selection efficiency for leptonic final states is
intentionally kept low. For the last three measurements (1996,
1997 and 1998), a 99.9\% pure hadronic sample was selected by requiring that the absolute value of the energy imbalance (ratio
of vector to scalar energy sum in the calorimeter) be
less than 0.6, that there be at least 22 GeV of visible
calorimetric energy, and that at least 4 charged tracks be reconstructed
in the central tracker; this selection was 92\% efficient. After
counting the number of hadronic decays for left- and right-handed
electron beams and forming a Left-Right Asymmetry, a small
experimental correction for
backgrounds (and negligible corrections for false asymmetries) was
applied; for the 1998 dataset this correction was 0.06\%. This
then yielded a value for the measured Left-Right Asymmetry (1998) of
\begin{eqnarray}
A_{LR}^{meas}& = & {1 \over P_e}
{N_L - N_R \over N_L + N_R} \\
& = & 0.1450 \pm 0.0030 \pm 0.0015
\label{eqn:alrmeas}
\end{eqnarray}
where the systematic uncertainty is dominated by the uncertainty in
the polarization scale. The translation of this result to the  
$Z^0$-pole asymmetry $A_{LR}^0$ was a $1.8 \pm 0.4$\% effect, where
the uncertainty arises from the precision of the center-of-mass energy
determination. This small error due to beam energy uncertainty is
slightly larger than seen previously (it has been quoted as $\pm
0.3$\%), and reflects the results of a scan of the Z peak used to
calibrate the energy spectrometers to LEP data, which was performed
for the first time during the 1998 run. This correction 
yields the 1998 preliminary result of
\begin{eqnarray}
  A_{LR}^0 & = &0.1487 \pm 0.0031 \pm 0.0017 \label{eqn:1998alr0}\\
  \sin^2\theta_W^{\sl eff}& =& 0.23130 \pm 0.00039 \pm 0.00022.
\label{eqn:1998sin2tw}
\end{eqnarray}

The six measurements $A_{LR}^0$ performed by SLD are shown in
Table \ref{tab:9298sin2tw}, along with their translations to 
$\sin^2\theta_W$. The values are consistent with each other yielding a
${\chi}^2$/$dof = 6.95$/$5$ for a straight line fit to the
(preliminary) average. 
The (preliminary) averaged results for 1992-98 are
\begin{eqnarray}
A_{LR}^0 = 0.1510 \pm 0.0025 
\label{eqn:9298alr0}\\
\sin^2\theta_W^{\sl eff} = 0.23101 \pm 0.00031
\label{eqn:9298sin2tw}
\end{eqnarray}

\section{Measuring \aelec , $A_{\mu}$ and $A_{\tau}$ Using Final State
Leptons}

Parity violation in the $Z^0$-lepton couplings is measured
by a maximum likelihood fit to the differential cross section
(Eqn. \ref{eqn:PolCross}) separately for the three leptonic final
states, including the effects of t-channel exchange for the $Z^0
\rightarrow e^+ e^-$ final state. For the 1996-98 data sample,
leptonic final states in the range $|\cos\theta| < 0.8$ are selected
by requiring that events have between two and eight charged tracks in
the CDC, at least one track with $p \ge 1$ GeV/c. One of the two event
hemispheres was required to have a net charge of -1, while the other one had to
have a net charge of +1. Events are identified as bhabha candidates if
the total calorimetric energy associated with the two
most energetic tracks is greater than 45 GeV. On the
other hand, if there is less than 10 GeV of associated energy
for each track, and there is a two-track combination with
an invariant mass of greater than 70 GeV/c$^2$, the event
is classified as a muon candidate. The selection criteria for tau
candidates is somewhat more complex. First, the $\tau^+\tau^-$ pair
invariant mass is required to be less than 70 GeV/c$^2$, 
with a separation angle between the two tau momentum vectors 
of at least 160$^{\circ}$. At least one track in the event must have a
momentum greater than 3 GeV/c. Each $\tau$ hemisphere must have an
invariant mass less that 1.8 GeV/$c^2$ (to suppress $Z^0$ hadronic
decays), and the associated energy in the LAC from each track must be
less than 27.5(20.0)~GeV for tracks with polar angles less than
(greater than) $|\cos\theta| = 0.7$. For the electron final state,
t-channel effects are incorporated into the angular
distribution, while for the tau final state, a
$\cos\theta$-dependent efficiency correction has been
applied to account for the correlation between visible
energy and net tau polarization.
Note that all channels provide information about $A_e$, which
comes in through the left-right cross section asymmetry.

The preliminary combined result for the 1993-1998 datasets are 
$A_{e}$=$0.1504 \pm 0.0072$, $A_{\mu}$=$0.120 \pm 0.019$ and
$A_{\tau}$=$0.142 \pm 0.019$. The result for $A_\mu$ (which is the
only direct measurement) reflects the best
single measurement, and is competitive with the overall LEP
value $A_{\mu}^{LEP} = 0.145 \pm 0.013$~\cite{LEPEWG}. A combination
of these values, assuming lepton universality, yields the value

\begin{eqnarray}
A_{lepton} & = & 0.1459 \pm 0.0063 \label{eqn:9398alepton}\\
\sin^2\theta_W^{\sl eff} & = & 0.2317 \pm 0.0008. 
\label{eqn:9398sin2twlepton}
\end{eqnarray}

\section{Measuring \abquark}
The determination of $A_b$ requires three ingredients: a measurement of the polarization, tag the $b$-quark (i.e., discriminate from $udsc$
flavored events) and tag the $b$-quark charge (i.e., was the underlying quark a
$b$ or $\bar{b}$?). The SLD has three separate analyses for obtaining
these quantities and measuring $A_b$: 1) jet charge, 2) cascade kaon
tag and 3) lepton tag. Although $A_b$ can be extracted from
Eqn. \ref{eqn:Alrfb}, most of these analyses use a maximum
likelihood treatment based on Eqn. \ref{eqn:PolCross}.

\begin{eqnarray}
L & = & {\left(1+{\cos}^{2}{\theta}\right)}{\left(1-A_{e}P_{e}\right)}+{2{\cos}{\theta}{\left(1-A_{e}P_{e}\right)}}
  \times {\left[
  A_{b}f_{b}{\left(1-2{\bar{\chi}}\right)}{\left(1-{\Delta}_{QCD}^{b}\right)}
  + A_{b}f_{b}{\left(1-{\Delta}_{QCD}^{c}\right)}
  + A_{bkg}f_{bkg} \right]}
\label{eqn:maxlike}
\end{eqnarray}

Here, $f_q$ is the fraction of tagged events of quark flavor $q$,
($1-2{\bar{\chi}}$) is a correction factor to account for asymmetry
dilution due to $B^{0}{\bar{B^{0}}}$ mixing and ${\Delta}_{QCD}^{q}$
is a cos$\theta$ dependent QCD correction for quark flavor $q$.

{\bf Momentum-Weighted Jet-Charge:} This analysis uses an inclusive
vertex mass tag to select a sample of $Z^{0} \rightarrow b{\bar{b}}$,
and the net momentum-weighted jet-charge to identify the sign of the
underlying quark. The track charge sum and difference between the two
hemispheres are used to extract the analyzing power from data, thereby
reducing MC dependencies and lowering many systematic effects.

{\bf Cascade Kaon:} The decay channel ${\bar{B}} \rightarrow D
\rightarrow K^-$ is exploited to tag the b charge in this analysis. The
gas-radiator data of the \v{C}erenkov Ring imaging Detector (CRID) is used
to identify charged kaons with high-impact parameter tracks. The charges of the
kaon candidates are summed in each hemisphere and the difference
between the two hemisphere charges is used to determine the polarity
of the thrust axis for the b-quark direction.

{\bf Leptonic Tag:} Electrons and muons are identified in hadronic
decays to enrich the $b{\bar{b}}$ event samples. The lepton charge
provides the quark-antiquark discrimination while the jet nearest in
direction to the lepton approximates the quark direction. The lepton
total and transverse momentum (with respect to the jet axis) are used
to classify each event by deriving probabilities for the decays $b
\rightarrow l^-$, ${\bar{b}} \rightarrow {\bar{c}} \rightarrow l^- $,
$b \rightarrow {\bar{c}} \rightarrow l^- $, ${\bar{c}} \rightarrow l^- $ and misidentified electrons.

\subsection{Preliminary Combined $A_b$}

The preliminary results for the three analyses are given in Table \ref{tab:Ab},
along with the corresponding statistical and systematic errors. The
jet-charge analysis uses the most amount of available data covering
1993-1998 partial data sample of approximately 400,000 $Z^{0}$ decays
(from a total of 550K events available). The lepton tag analysis uses
data from 1993-1996, or approximately 200,000 events. Finally, the
kaon tag uses the 1993-1995 runs, or approximately 150k events. The
SLD average is also given in the table along with the LEP average for
comparison.

\section{Summary of (Preliminary) SLD Results}

Combining $A_{LR}$ and $A_{leptons}$ obtains the preliminary SLD
result for the effective weak mixing angle,
\begin{eqnarray}
{\sin}^2{\theta}_{W}^{\sl eff} & = & 0.23109 \pm 0.00029. 
\label{eqn:combsin2tw}
\end{eqnarray}

Combining this result with LEP gives the SLD-LEP world average,
\begin{eqnarray}
{\sin}^2{\theta}_{W}^{\sl eff} & = & 0.23155 \pm 0.00018. 
\label{eqn:combsldlepsin2tw}
\end{eqnarray}

These results are shown in Figure 1. 
It is interesting to note that the world values for
$\sin^2\theta_W^{\sl eff}$ from lepton measurements (the results
presented here, the LEP leptonic averages $A_{FB}^l$ and $A_e$ , and
the LEP results of $A_\tau$ from $\tau$-polarization) are in excellent
agreement with each other, and differ by $\sim2.3\sigma$ from
those obtained from ``hadron-only'' measurements ($A_{FB}^b$,
$A_{FB}^c$, and $Q_{FB}$).

The combined preliminary SLD result for $A_b$ is,
\begin{eqnarray}
A_{b} & = & 0.866 \pm 0.036. 
\label{eqn:abresult}
\end{eqnarray}

These results for $A_b$ and $A_{LR}$ have been transformed and
displayed in Figure 2, following the scheme of a full
$Z^{0}b{\bar{b}}$ coupling analysis proposed by Takeuchi et
al. \cite{takeuchi} Here, ${\zeta}_b$ is a variable which isolates
deviations in the right-handed $b$-quark neutral current couplings. This is plotted versus
${\delta}{\sin}^2{\theta}_{W}^{\sl eff}$. Since LEP's $A^b_{FB}$ is a
mixture of $A_b$ and $A_e$, it comes in at $45^\circ$. The standard
model is the horizontal line below the 68\% and 90\% CL contours.



\begin{figure}[hbt]
\centerline{\epsfxsize 6 truein \epsfbox{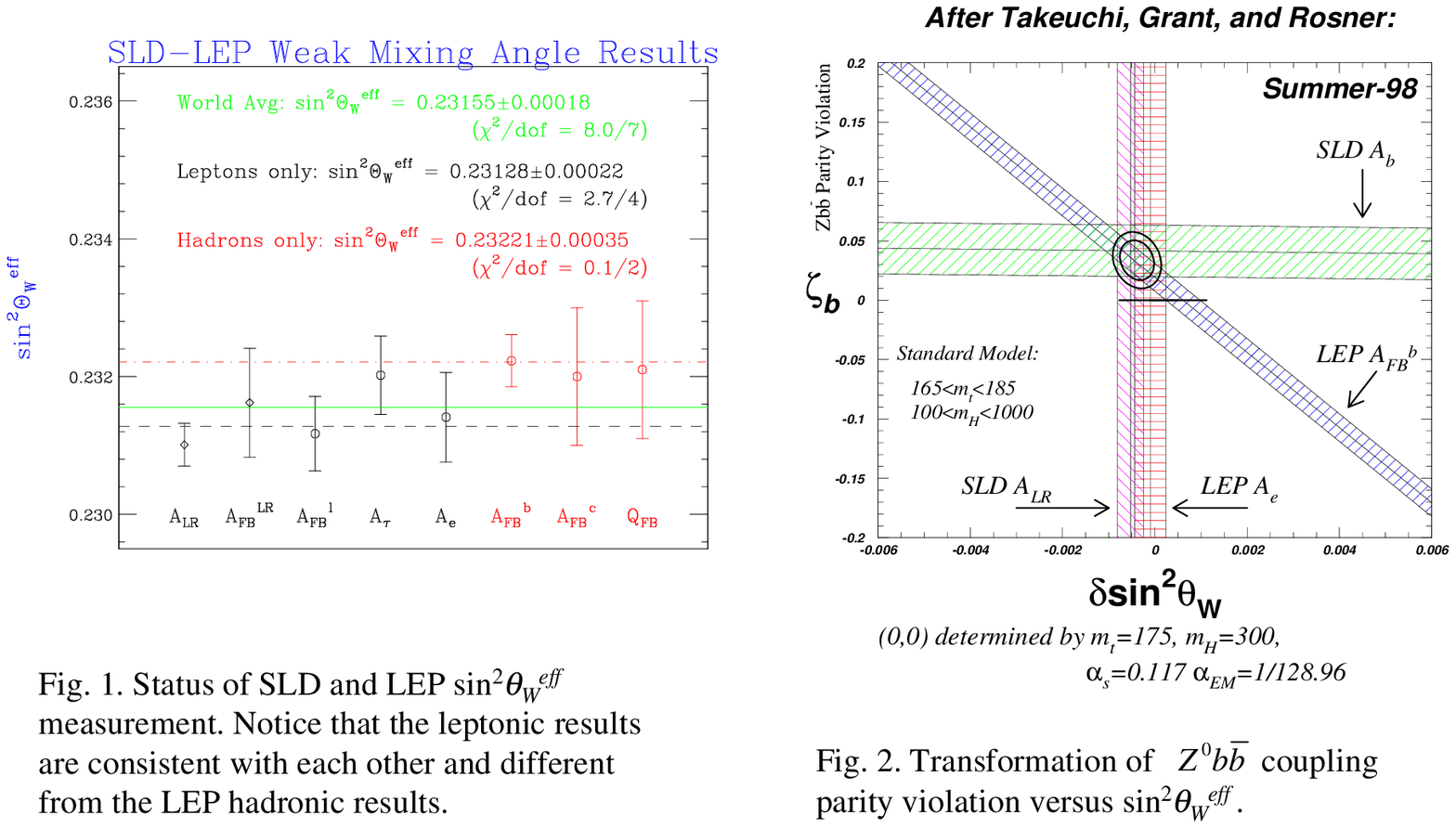} }   
\vskip 0 cm
\end{figure}

\begin{table}[htb] 
\caption{The six measurements of $A_{LR}^0$ and $\sin^2\theta_W^{\sl eff}$ 
performed by the SLD  experiment. The *'s indicate preliminary results.}
\label{tab:9298sin2tw}
\begin{tabular}{lllll}
Run & $A_{LR}^{0}$ & $\hspace{0.0in}\delta A_{LR}^{0}$ &
$\sin^{2}\theta_{w}^{\mbox{\tiny eff}}$ &$\hspace{0.0in}\delta \sin^{2}\theta_{w}^{\mbox{\tiny eff}} $\\
\tableline
1992 & 0.100 & $\pm 0.044 \pm 0.004 $ & 0.2378 & $\pm 0.0056 \pm 0.0005$ \\
1993 & 0.1656 & $\pm 0.0071\pm 0.0028 $ & 0.2292 & $\pm 0.0009 \pm 0.0004$ \\
1994-95 & 0.1512 & $\pm 0.0042 \pm 0.0011$ & 0.23100  & $\pm 0.00054 \pm 0.00014$ \\
1996 & $0.1570^*$ & $\pm 0.0057\pm 0.0011^*$ & $0.23025^*$ & $\pm 0.00073\pm 0.00014^*$ \\ 
1997 & $0.1475^*$ & $\pm 0.0042\pm 0.0016^*$ & $0.23146^*$ & $\pm 0.00054\pm 0.00020^*$ \\ 
1998 & $0.1487^*$ & $\pm 0.0031\pm 0.0017^*$ & $0.23130^*$ & $\pm 0.00039\pm 0.00022^*$ \\
\tableline
Combined & $0.1510^*$ & $\pm 0.0025^*$ & $0.23101^*$ & $\pm 0.00031^*$ \\ 
\end{tabular}
\end{table}

\begin{table}[htb]
\caption{Combined results of $A_b$ from the 1993-98 data(partial
sample). The SLD results are preliminary and are thru Summer 1998. The
LEP results are included for comparison.  Note that the LEP
measurement are derived from $A_b = 4A_{FB}^{0,b}$/$\left( 3A_e
\right)$, where $A_e = 0.1491 \pm 0.0018$ (the combined SLD $A_{LR}$
and LEP $A_{lepton}$. }
\label{tab:Ab}
\begin{tabular}{lcc}
 Analysis & $A_{b} \pm \left( stat \right) \pm \left( syst \right)$  &
Combined \\
\tableline
SLD Jet-Charge & $0.849 \pm 0.026 \pm 0.031$ & \\
SLD Lepton & $0.932 \pm 0.058 \pm 0.038$ & \\
SLD Kaon & $0.854 \pm 0.088 \pm 0.106$ & \\
SLD Average & & $0.866 \pm 0.036$ \\
\tableline
ALEPH Lept & $0.908 \pm 0.041 \pm 0.020$ & \\
DELPHI Lept & $0.904 \pm 0.057 \pm 0.026$ & \\
L3 Lept & $0.869 \pm 0.055 \pm 0.030$ & \\
OPAL Lept & $0.851 \pm 0.038 \pm 0.020$ & \\
ALEPH Jet-Charge & $0.953 \pm 0.037 \pm 0.029$ & \\
DELPHI Jet-Charge & $0.898 \pm 0.042 \pm 0.021$ & \\
L3 Jet-Charge & $0.806 \pm 0.106 \pm 0.051$ & \\
OPAL Jet-Charge & $0.898 \pm 0.047 \pm 0.037$ & \\
LEP Average & & $0.885 \pm 0.022$ \\
\tableline
Standard Model & & $0.935$ \\
\end{tabular}
\end{table}

\end{document}